\documentclass[reprint,amsmath,amssymb,aps]{revtex4-2}
\usepackage{graphicx,subfigure,bm,amssymb,amsmath,hyperref,dcolumn}
\usepackage{color,multirow,supertabular,float}
\usepackage{mathrsfs}
\usepackage{polyglossia}
\setdefaultlanguage{english}
\usepackage{physics}
\usepackage{amsmath}
\usepackage{tikz}
\usepackage{mathdots}
\usepackage{yhmath}
\usepackage{cancel}
\usepackage{color}
\usepackage{siunitx}
\usepackage{array}
\usepackage{multirow}
\usepackage{amssymb}
\usepackage{gensymb}
\usepackage{tabularx}
\usepackage{extarrows}
\usepackage{booktabs}
\usetikzlibrary{fadings}
\usetikzlibrary{patterns}
\usetikzlibrary{shadows.blur}
\usetikzlibrary{shapes}
\usepackage[compat=1.1.0]{tikz-feynman}

\begin{abstract}
---
\end{abstract}

\begin{document}
\title{The 4-$\epsilon$ Expansion for Long-range Interacting Systems}
\author{Zhiyi Li$^{1,2}$}
	\author{Kun Chen$^{3}$}
        \email{chenkun@itp.ac.cn}
	\author{Youjin Deng$^{1,2,4}$}
        \email{yjdeng@ustc.edu.cn}
	\affiliation{$^{1}$ Department of Modern Physics, University of Science and Technology of China, Hefei, Anhui 230026, China}
	\affiliation{$^{2}$ Hefei National Laboratory, University of Science and Technology of China, Hefei 230088, China}
        \affiliation{$^{3}$CAS Key Laboratory of Theoretical Physics, Institute of Theoretical Physics, Chinese Academy of Sciences, Beijing 100190, China}        \affiliation{$^{4}$ Hefei National Research Center for Physical Sciences at the Microscale and School of Physical Sciences, University of Science and Technology of China, Hefei 230026, China}

\begin{abstract}
The establishment of the Wilson-Fisher fixed point (WFP) for $O(n)$ spin models in $d=4-\epsilon$ dimensions stands as a cornerstone of the renormalization group (RG) theory for critical phenomena. However, when long-range (LR) interactions, algebraically decaying as 
$\propto 1/r^{d+\sigma}$, are introduced, 
the fate of the short-range WFP (SR-WFP) has remained a subject of intense debate since the 1970s.
We employ two complementary techniques—the standard field-theoretic RG and a perturbative bootstrap scheme, and perform the $\epsilon$-expansion calculations up to the two-loop level.
We show that, as long as $\sigma<2$, the SR-WFP becomes unstable and a stable LR-WFP emerges,
and, in the non-classical regime with $d/2 < \sigma < 2$, 
the critical exponents, including the anomalous dimension,
are functions of $\epsilon$, $\delta=2-\sigma$ and $n$,
which reduce to the exact results in the limiting cases $\epsilon \to 0$, $\delta \to 0$ or $n \to \infty$.
Our $(4-\epsilon)$-expansion calculations support the scenario that 
the threshold between the LR- and SR-WFP occurs strictly at $\sigma_*=2$, 
well consistent with the recent high-precision numerical study
while different from the widely accepted Sak's criterion.
\end{abstract}
\maketitle

\section{Introduction}
The principle of universality is a cornerstone of statistical physics, positing that the critical behavior of diverse physical systems depends only on global symmetries and dimensionality, rather than on microscopic details. However, this paradigm faces a profound challenge when interactions are non-local. Long-range (LR) interactions, characterized by a power-law decay $J(r) \sim 1/r^{d+\sigma}$, are ubiquitous in nature, governing phenomena from dipolar ferromagnetism and screened Coulomb interactions to the dynamics of biophysical networks~\cite{LRbook,spivak2004,lahaye2009,peter2012}. In recent years, the study of such systems has gained renewed urgency due to the rapid development of quantum simulators. Experimental platforms based on trapped ions~\cite{Monroe2021, Britton2012,lewis2023} and Rydberg atom arrays~\cite{Browaeys2020, Saffman2010,schauss2012} can now engineer tunable long-range couplings, allowing for the direct exploration of critical universality classes that were previously theoretical abstractions.

The theoretical framework for these systems is the long-range O($n$) spin model. Despite decades of study, the nature of the crossover from long-range to short-range (SR) universality remains one of the most enduring controversies in the field~\cite{defenu2023}. The central question concerns the threshold decay exponent $\sigma_*$ at which the non-local interaction ceases to dominate the critical fluctuations.
Specifically, the model is defined by the Hamiltonian:
\begin{equation}
\mathcal{H} = -J \sum_{i<j} \frac{1}{r_{ij}^{d+\sigma}} \mathbf{S}_i \cdot \mathbf{S}_j - \sum_i \mathbf{H}_i \cdot \mathbf{S}_i ,
\label{eq:hamiltonian}
\end{equation}
where $\mathbf{S}_i$ is an $n$-component unit vector on a $d$-dimensional lattice, $\mathbf{H}_i$ refers to the external field and $J>0$ is the coupling constant.  In the continuum limit, the competition between local and non-local fluctuations is encoded in the Ginzburg-Landau-Wilson effective action:
\begin{align}
\mathcal{S} =&\int d^d x  \left[\frac{1}{2} \phi(x) (r_0 +K_L(-\Delta)^{\sigma/2} + K_s( -\Delta)) \phi(x) \right. \notag\\
&\left .+ u_0(\phi(x)^2)^2 + h \phi(x)\right ], 
\label{eq:action}
\end{align}
where the fractional Laplacian $(-\Delta)^{\sigma/2}$ represents the non-analytic LR kinetic term (corresponding to $k^\sigma$ in the momentum space) and $-\Delta$ represents the analytic SR term ($k^2$) generated by lattice regularization~\cite{nijboer1957calculation}. 

Historically, the renormalization group (RG) flow of this action has been described by two conflicting scenarios. Following 
the celebrated work for dimension $d=4-\epsilon$ in local $\phi^4$ theory~\cite{PhysRevLett.28.240}, Fisher, Ma, and Nickel~\cite{Fisher1972} employed an expansion with $\epsilon' = 2\sigma - d$ 
and obtained a non-trivial LR fixed point, which we shall 
call ``LR Wilson-Fisher fixed point" (LR-WFP).
By fixing the decay exponent $\sigma<2$ and expanding 
in the spatial dimension $d < 2 \sigma$ around the LR Gaussian fixed point (LR-GFP) at $d=2\sigma$, they had the anomalous dimension $\eta$ and the correlation-length exponent $\nu$, associated with LR-WFP, 
\begin{align}
\eta &= 2 - \sigma + O(\epsilon'^3), \label{eq:eta_expand0} \\ 
\quad \frac{1}{\nu} &= \sigma - \frac{(n+2)}{(n+8)} \epsilon' + O(\epsilon'^2).
\label{eq:eta_expand1}
\end{align}
The exponent $\eta$ sticks at the mean field value $2-\sigma$ up to $O(\epsilon'^2)$ order, 
and it was further suggested that it might be true to all orders~\cite{Fisher1972}.


This scenario predicted that the $k^\sigma$ term dominates the infrared physics whenever $\sigma < 2$, placing the crossover strictly at $\sigma_* = 2$. However, this scenario implied a suspicious discontinuity in the anomalous dimension $\eta$, which would jump abruptly from its long-range mean-field value $\eta_{\text{LR}} = 2-\sigma$ to the short-range value $\eta_{\rm SR}$ at the boundary.


To resolve this discontinuity, Sak \cite{Sak1973} proposed a refined criterion that has since become the standard paradigm. He argued that the LR-WFP remains stable only when the scaling dimension of the long-range interaction term dominates over that of the short-range fluctuations generated by renormalization. This leads to the prediction $\sigma_* = 2 - \eta_{\text{SR}}$, generally known as Sak's criterion. Under this scenario, the critical exponents vary continuously, satisfying $\eta = \max(2-\sigma, \eta_{\rm SR})$, thereby ensuring a smooth crossover between regimes. 

Despite decades of study, the critical behavior of long-range systems remains a vibrant frontier in statistical physics, capturing sustained attention across theoretical and numerical communities. For years, Sak's criterion was supported by a consensus of numerical and theoretical studies. 
Numerically, early Monte Carlo studies~\cite{Luijten2002} and finite-size scaling analyses~\cite{angelini_relations_2014} argued that deviations from Sak's prediction were merely subdominant corrections rather than intrinsic physics. Theoretically, detailed RG analyses supporting Sak's idea were performed, which treated the long-range term perturbatively and assumed it should not be renormalized~\cite{honkonen_crossover_1989}. Later on, high-order perturbative RG calculations refined other critical exponents in the $\sigma<\sigma_*$ regime, assuming the standard expansion remained valid~\cite{benedetti_long-range_2020}, while recent advances in conformal field theory (CFT) and the conformal bootstrap \cite{paulos2016,behan2017,Behan_2024}, provided a rigorous non-perturbative framework for non-local CFTs. These studies described the long-range Ising model as a defect theory or via holographic realizations, generally supporting a scenario of continuous variation of critical data consistent with the continuity assumption inherent in Sak's criterion.

However, the consensus has fractured. Hints of this breakdown appear in early field-theoretic analyses by Yamazaki~\cite{yamazaki_department_1977}, who demonstrated that the renormalization group flow becomes unstable in the intermediate regime where interaction ranges compete, foreshadowing the failure of the standard crossover description. Moreover, recent large-scale simulations through enhanced cluster algorithm of the 2D Ising ($n=1$)~\cite{xiao_saks_2026,picco_critical_2012,blanchard2013}, XY ($n=2$)~\cite{Xiao2025_XY,yaodingyun2025}, and Heisenberg ($n=3$)~\cite{yao_nonclassical_2025} models, as well as percolation (within the $\phi^3$ field-theoretical description)~\cite{liu_two-dimensional_2025,Grassberger_2013}, 
consistently reveal a sharp universality change at $\sigma_* = 2$. Specifically, by targeting geometric observables like the Fortuin-Kasteleyn critical polynomial—which acts as a highly sensitive probe of conformal data distinct from standard magnetic order parameters—these investigations provided unambiguous evidence that the universality class changes strictly at $\sigma_* = 2$.
The Goldstone-mode properties in the low-temperature ordered phase 
also display a crossover at $\sigma_*=2$, consistent with the insights 
from L\'evy flights--i.e., long-range simple random walk~\cite{janssen1999levy}.   

These high-precision and systematic numerical studies suggest that
Sak's criterion might exhibit a subtle but intrinsic limitation.
Sak's criterion is based on two key conjectures that
the anomalous dimension is locked at the mean-field value $\eta=2-\sigma$ 
to all orders in $\epsilon'$, and the SR-WFP must remain stable—and thus govern the critical behavior—throughout the extended range $\sigma \in (2-\eta_{\rm SR}, 2]$.
The second conjecture also implies that the LR-WFP becomes unstable in the range 
$\sigma \in (2-\eta_{\rm SR}, 2]$.
However, to our knowledge, no solid theoretical justification has been given 
for any of these assumptions after more than 50 years' investigation.

In this paper, we reexamine the fate of the SR-WFP under long-range interactions and resolve the controversy regarding the instability of the SR-WFP for $\sigma \leq 2$. To achieve this, we employ a systematic renormalization group analysis controlled by the small parameter $\epsilon = 4-d$. Our analysis explicitly targets the entire non-mean-field regime $\sigma > d/2$.
Crucially, the non-classical domain, with $d/2 < \sigma < 2$, falls strictly within the validity scope of the $\epsilon$-expansion, ensuring that our perturbative framework remains well-defined. This distinguishes our approach from the standard $\epsilon' = 2\sigma - d$ expansion, which is fundamentally limited because it expands strictly from the mean-field line $d=2\sigma$ along the direction of $d$ or $\sigma$, thereby focusing solely on the physics near the mean-field boundary and failing to capture the physics as $\sigma \to 2$.

We demonstrate the validity of our expansion by applying it within two complementary RG schemes: the standard field-theoretical RG expanding around the LR Gaussian fixed point (LR-GFP), and a perturbative bootstrap scheme which allows one to expand the theory of the Wilson-Fisher-like free field. Our results reveal that the pole structure contributes to the renormalization flow in a non-trivial manner, generating two-loop corrections to the critical exponent $\eta$, explicitly refuting the first conjecture of Sak. Consequently, by systematically probing the physics that significantly deviates from the traditional mean-field regions, 
including the strong LR case $\sigma \leq d/2$ and the spherical model ($n \to \infty$),
our results successfully interpolate between the mean-field limit and the strict short-range limit ($\sigma=2$). We derive the corrected scaling relations and demonstrate that the LR-WFP remains stable and distinct from the SR-WFP up to the threshold $\sigma_*=2$. This theoretical foundation not only explains the recent numerical findings but also establishes a unified framework connecting statistical mechanics to non-local field theories such as Lévy flights, fractional quantum mechanics, and turbulence~\cite{Metzler2000, Laskin2000,adzhemyanbook}.

\section{Main Results}
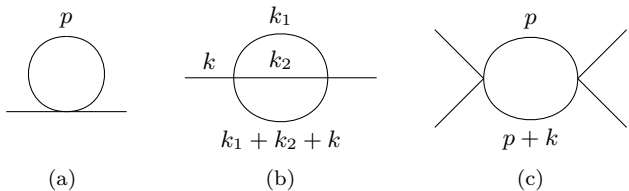
\begin{figure}
    \centering
    \subfigure[]{ \begin{tikzpicture}
  \begin{feynman}
    \vertex (i);
    \vertex [right=0.8cm of i] (v);        
    \vertex [above=1.0 cm of v, label =\(p\)] (loop_top); 
    \vertex [right=0.8 cm of v] (o);
    \vertex [below = 0.6cm of v ] (x);

    \diagram* {
      (v) -- [color = white] (x),
      (i) -- (v),
      (v) -- [edge label = \(~~~\) ](o),
      (v) -- [ half left, looseness=1.73] (loop_top),
      (loop_top) -- [ half left, looseness=1.73] (v),
      
    };
  \end{feynman}
\end{tikzpicture}\label{fig:diag-tadpole}}
    ~~~~    
    \subfigure[]{\begin{tikzpicture}
  \begin{feynman}
    \vertex (i1);
    \vertex [right=0.65cm of i1] (a);
    \vertex [right= 1.25cm of a] (b);
    \vertex [right=0.65cm of b] (f1);

    \diagram* {
      (i1) --[edge label=\(k\)] (a),                  
      (a) -- [half left, looseness=1.6, edge label =\(k_1\) ] (b),       
      (b) -- [half left, looseness=1.6,edge label=\(k_1+k_2+k\) ] (a),       
      (a) -- [edge label=\(k_2\)] (b),
      (b) -- (f1),
    };
  \end{feynman}
\end{tikzpicture}\label{fig:diag-sunset}}
    ~~~~
    \subfigure[]{\begin{tikzpicture}
  \begin{feynman}
    \vertex (a);
     \vertex [left=0.65cm of a, yshift=0.65cm] (i1);
    \vertex [left=0.65cm of a, yshift=-0.65cm] (i2);
    \vertex [right=1.25cm of a] (b); 
    \vertex [right=0.65cm of b, yshift=0.65cm] (f1);
    \vertex [right=0.65cm of b, yshift=-0.65cm] (f2);

    \diagram* {
      (i1) --  (a),
      (i2) -- (a),
      (a) -- [half left,  edge label = \(p\)] (b),
      (b) -- [half left, edge label = \(p+k\)] (a),
      (b) -- (f1),
      (b) -- (f2),
    };
  \end{feynman}
\end{tikzpicture}\label{fig:diag-bubble}}
    \caption{Feynman diagrams used in the perturbative expansion: (a) one-loop self-energy tadpole, (b) two-loop sunset self-energy, and (c) one-loop interaction bubble. }
    \label{fig:diagrams}
\end{figure}

By defining the dimensionless renormalized coupling $u$ and mass $r$ at the energy scale $\mu$, and employing the $d = 4-\epsilon$ expansion within two complementary renormalization group schemes—the standard field-theoretical RG and the perturbative bootstrap scheme—we analyze the Feynman diagram shown in Fig.~\ref{fig:diagrams} up to the two-loop level and obtain the LR-WFP coordinates:
\begin{align}
u^* &= \frac{2\pi^2}{n+8} (\epsilon-2\delta)+O(\epsilon^2), 
\nonumber \\
r^* &= 0 + O(\epsilon^2) , 
\end{align}
and derive the associated critical exponents as 
\begin{align}
    \eta &= \delta + \frac{n+2}{(n+8)^2} \frac{(\epsilon-2\delta)^3}{2\epsilon -3\delta} +O(\epsilon^3), 
\label{eq:double_expand_eta} \\
\frac{1}{\nu} & = 2-\delta -\frac{n+2}{n+8}(\epsilon-2\delta) + O(\epsilon^2) .
\label{eq:double_expand_nu}
\end{align}
Here, the parameter $\delta = 2 - \sigma$ satisfies $0 < \delta < \epsilon/2$
in the non-classical regime ($d/2 < \sigma <2$),
and, by definition, is a small variable in the $\epsilon$-perturbative results.
Moreover, it can be calculated that
the RG exponent for the interaction term in Eq.~(\ref{eq:action}) 
is now irrelevant as $y_u  = -(\epsilon - 2\delta) +O(\epsilon^2)$,
confirming the renormalizability up to the two-loop level.

We emphasize that Eqs.~(\ref{eq:double_expand_eta}) and~(\ref{eq:double_expand_nu}) apply only in the case $d=4-\epsilon$ within the non-classical regime as $0< \delta < \epsilon/2$. 
Nevertheless, for the SR case with $\sigma > 2$, 
we can set $\delta=0$ and obtain the celebrated $(4-\epsilon)$-perturbative results  
for the short-range and low-dimensional (SR-LD) case,
having $\sigma >2$ and $2 < d <4$,
\begin{align}
\eta &= \frac{n+2}{2(n+8)^2} \epsilon^2 +O(\epsilon^3), \\
\frac{1}{\nu} & = 2 -\frac{n+2}{n+8} \epsilon + O(\epsilon^2) .
\label{eq:double_expand_nu1}
\end{align}
Intriguingly, there is a cancellation of $\epsilon$ in the numerator and denominator of Eq.~(\ref{eq:double_expand_eta}), leading to $\eta \approx O(\epsilon^2)$. 
For the SR and high-dimensional (SR-HD) case $(\sigma>2, d> 4)$, 
the SR-GFP results, $\eta=0, \nu=1/2$, are obtained by setting $\epsilon=0$.

Another interesting limit is to take the $n \to \infty$ limit
to obtain the spherical model~\cite{Joyce1966}. The known exact results,  $\eta = 0$ 
and $1/\nu = d-2$ for the SR case ($\sigma >2$) and $\eta = 2-\sigma $ and $1/\nu = d-\sigma$ 
for the LR case ($ d/2 < \sigma < 2$), can be straightforwardly reduced 
from Eqs.~(\ref{eq:double_expand_eta}) and~(\ref{eq:double_expand_nu}).

For the non-classical regime with a fixed value of $d/2< \sigma<2$, 
we note $\epsilon' = 2 \sigma-d = \epsilon - 2\delta$, 
and, by rewriting Eqs.~(\ref{eq:double_expand_eta}) and~(\ref{eq:double_expand_nu}) explicitly in terms of $\epsilon'$
and $\delta$, obtain 
\begin{align}
\eta &= \delta + \frac{n+2}{(n+8)^2} \frac{\epsilon'^3}{2\epsilon' +\delta} +O(\epsilon'^4), 
\label{eq:double_expand_eta2} \\
\frac{1}{\nu} & = 2-\delta -\frac{n+2}{n+8} \epsilon' + O(\epsilon'^2),
\label{eq:double_expand_nu2}
\end{align} 
where the result for $1/\nu$ reduces to Eq.~(\ref{eq:eta_expand1}).
Interestingly, the value of $\eta$ is now of $O(\epsilon'^3)$, 
without introducing inconsistency with Eq.~(\ref{eq:eta_expand0}). 
This suggests that an alternative way to obtain Eqs.~(\ref{eq:double_expand_eta}) 
and~(\ref{eq:double_expand_nu}) could be to perform the $\epsilon'$ expansion up to 
higher orders, an important but challenging task to be done in the future.

\section{Dimensional Analysis and Sketch of RG Flows}  
To investigate the critical behavior and the RG flow of the model, we begin with a dimensional analysis of the effective action. The competition between the long-range and short-range interactions is dictated by the relative relevance of the nonlocal term $K_L(-\Delta)^{\sigma/2}$ versus the local gradient term $K_s(-\Delta)$ in Eq.~\eqref{eq:action}. 

Under a coarse-graining transformation with scale factor $b$, 
the spatial coordinate and the momentum are rescaled as $x'=x/b$ and $p' = b p$, respectively.
By setting the dimension of the momentum to be unity as $[p]=1$, 
we then have $[x]=-1$ and $[d^d x]=-d$. Since the effective action $\mathcal{S}$ appears in 
the Boltzmann distribution as $e^{-\mathcal{S}}$, 
its dimension is  zero, $[\mathcal{S}]=0$, and so is for each term in $\mathcal{S}$.

\begin{figure}
    \centering
    \includegraphics[width=0.95\linewidth]{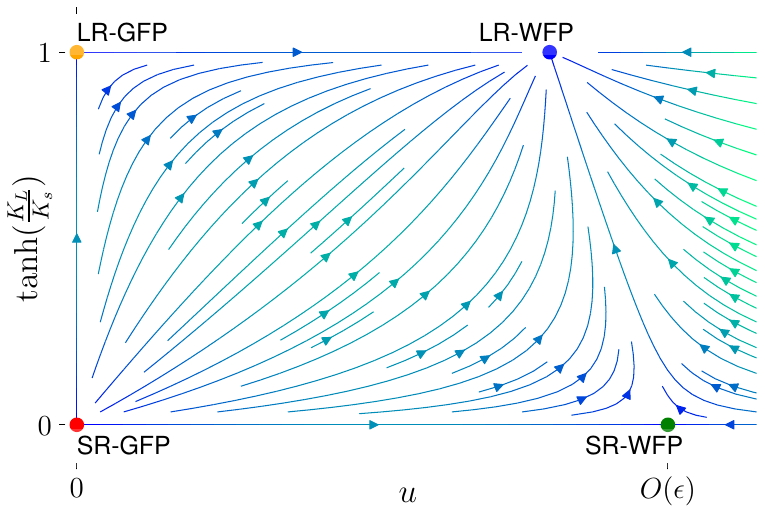}
    \caption{Sketch of RG flows for the $O(n)$ spin model in the LR-LD regime with $\sigma < 2$ and $\sigma < d < 2\sigma$.
    In the $(4-\epsilon)$ dimensions, the SR-WFP and the LR-WFP are located at the dimensionless 
    interaction $u \sim O(\epsilon)$. 
    }
    \label{fig:RG-flow}
\end{figure}

Let us now analyze the canonical dimensions for the SR-GFP, at which the amplitude $K_s$ 
in the SR gradient term remains unchanged under the coarse-graining transformation--i.e., $[K_s]=0$.
From $[ \int \Delta \phi^2 d^d x]=0$, we have $[\Delta\phi^2]=d$ and $[\phi]=(d-2)/2$.
The analysis of the mass, interaction, and LR gradient terms 
gives $[r_0]=2$, $[u_0]=4-d$ and $[K_L]=2-\sigma$, respectively.
From this tree-level analysis, the critical exponents describing the RG flow around SR-GFP are given by
\begin{equation}
y_t = 2,\ \   y_{K_L} = 2-\sigma, \ \  y_u=4 -d ,
\label{eq:exponent_SR-GFP}
\end{equation}
suggesting that the SR-GFP is stable if and only if $\sigma>2$ and $d>4$.

Analogous analysis can be applied to the LR-GFP that requires $[K_L]=0$ for the LR gradient term,
giving $[\phi]=(d-\sigma)/2$, $[r_0]=\sigma$, $[u_0]=2\sigma -d$ and $[K_S]=\sigma-2$.
The LR-GFP is thus stable when $\sigma<2$ and $d>2\sigma$, and the associated 
critical exponents are
\begin{equation}
 y_t= \sigma,\ \   y_{K_S}= \sigma -2, \ \  y_u= 2\sigma -d.
\label{eq:exponent_LR-GFP}
\end{equation}

The critical exponents, in Eqs.~(\ref{eq:exponent_SR-GFP}) and~(\ref{eq:exponent_LR-GFP}),
immediately imply a number of threshold values for $d$ and $\sigma$. 
From the RG exponent $y_u$ for interactions, one obtains the upper spatial dimensionality $d_{\rm up}=4$ for the SR-GFP and $d_{\rm up}=2\sigma$ for the LR-GFP, giving  $d_{\rm up}= \min(2\sigma, 4)$.
Also, note that the thermal RG exponent $y_t$ has to be bounded by $y_t \leq d$, 
which defines the lower spatial dimensionality  $d_\ell =\min (\sigma,2)= d_{\rm up}/2$
such that, for $d < d_\ell$, the $O(n)$ spin model cannot exhibit finite-temperature phase transitions.
Thus, the systems can be classified to be in the high-dimensional (HD) regime for $d> d_{\rm up}$, 
be in the low-dimensional (LD) regime for $d_\ell < d< d_{\rm up}$, and
be in zero-$T_c$ regime for $d<d_\ell$. 
From the effective RG exponents $y_{K_L}$ and $y_{K_S}$, one has a threshold $\sigma_*=2$ 
for the decay exponent $\sigma$, so that the systems are called to be long-ranged (LR) 
for $\sigma<2$ and short-ranged (SR) for $\sigma>2$. 

More importantly, in the LD regime ($d_\ell < d< d_{\rm up}$) where 
the Gaussian fixed points become unstable and new stable non-trivial fixed points,
called SR-WFP for $\sigma>2$ and LR-WFP for $\sigma<2$, emerge, 
our $\epsilon$-perturbative results, Eqs.~(\ref{eq:double_expand_eta})
and~(\ref{eq:double_expand_nu}), strongly indicate that 
the boundary for the stability of the SR-WFP and the LR-WFP remains to be at  $\sigma_*=2$.

As a consequence, one can partition the $(\sigma, d)$ parameter plane into 
five regimes: SR-HD ($\sigma >2, d>d_{\rm up}$), SR-LD ($\sigma >2, d_\ell < d< d_{\rm up}$), 
LR-HD ($\sigma < 2, d>d_{\rm up}$), LR-LD ($\sigma < 2, d_\ell < d< d_{\rm up}$), 
and zero-$T_c$ regime ($d < d_\ell$), as depicted in Ref.~\cite{xiao2026universality}.
The boundaries between different regimes are simply straight lines,
and the point $(\sigma=2,d=2)$ acts as a multi-regime threshold 
connecting the zero-$T_c$, SR-LD and LR-LD regimes,
illustrating a simple and beautiful geometric feature.
In contrast, within the framework of Sak's criterion, 
it is not clear how to depict the boundaries around $(\sigma=2,d=2)$.

From the analyses about the stability for the SR-GFP, LR-GFP, SR-WFP, and LR-WFP,
the RG flows can be qualitatively sketched for each of the five regimes.
For the LR-LD regime $(d/2< \sigma <2, d_\ell < d < d_{\rm up})$, 
which is non-classical and of most interest, 
Fig.~\ref{fig:RG-flow} depicts the RG flows in the parameter plane of $u$ 
and $\tanh (K_L/K_S)$, where $u$ is for the dimensionless interaction strength 
and $\tanh (K_L/K_S)$ represents the dimensionless relative strength between the LR 
and SR kinetic terms. 

In Fig.~\ref{fig:RG-flow}, for the SR-GFP and the LR-GFP the RG exponents can be readily read out from Eqs.~(\ref{eq:exponent_SR-GFP}) and~(\ref{eq:exponent_LR-GFP}).  
Near the SR-WFP and the LR-WFP, the interaction RG exponent is also available 
in the $(4-\epsilon)$ dimensions, which is $y_u=-\epsilon$ for the SR-WFP 
and $y_u=-(\epsilon-2\delta)$ for the LR-WFP.

Next, we discuss the RG exponents for the SR-WFP and the LR-WFP in the direction of $\tanh(K_L/K_S)$.
We note that, at the Gaussian fixed points, there are two free propagators $G_{0,2}^{-1} = k^{2}$ and $G_{0,\sigma}^{-1} = k^{\sigma}$. 
Also, an alternative way to obtain the dimension of $K_L$ at the SR-GFP  
is $[K_L]=[G_{0,2}^{-1}]-[G_{0,\sigma}^{-1}]=2-\sigma$, and, 
similarly, $[K_s]$ at the LR-GFP is $[K_S]=[G_{0,\sigma}^{-1}]-[G_{0,2}^{-1}]=\sigma-2$.
At the SR-WFP with $\sigma>2$ and $d<4$, 
the dominant free propagator $G_{0,2}^{-1}=k^2$ is renormalized to be 
the bold one $G_{2}^{-1}=k^{2-\eta_{\rm SR}}$,
and we expect that the subdominant free propagator $G_{0,\sigma}^{-1}=k^\sigma$ 
is renormalized, in a similar way, to be $G_{\sigma}^{-1}=k^{(2-\eta_{\rm SR})(\sigma/2)}$.
Thus, we conjecture that the dimension of the LR gradient amplitude becomes
$[K_L]= [G_{2}^{-1}]-[G_{\sigma}^{-1}]=(2-\eta_{\rm SR}) (2-\sigma)/2>0$. From this point of view, the RG flows around the SR-WFP can be approximately described by 
\begin{equation}
    y_{K_L} = \frac{1}{2}(2-\eta_{\rm SR}) (2-\sigma).
\end{equation}
Analogously, at the LR-WFP, we expect to have two effective bold propagators 
as $G_{\sigma}^{-1}=k^{2-\eta_{\rm LR}}$ and $G_{2 }^{-1}=k^{(2-\eta_{\rm LR})\sigma/2}$,
giving $[K_s] = [G_{\sigma}^{-1}]-[G_{2}^{-1}]= (2-\eta_{\rm LR}) (\sigma-2)/\sigma <0$, which suggests that
\begin{equation}
    y_{K_S} = \frac{1}{\sigma}(2-\eta_{\rm LR}) (\sigma-2).
\end{equation}

\section{Methods} Having analyzed the global flow topology and stability of the fixed points, we now turn to the rigorous renormalization group formalism employed to derive the quantitative critical exponents. As shown by our previous discussion within tree-level analysis, the short-range term is irrelevant at $\sigma<2$, therefore, we consider the bare field only with the long-range kinetic term as $K_s = 0$ and without external field as $h=0$ in Eq.~\eqref{eq:action}. 

\subsection{Perturbative bootstrap scheme}
Standard renormalization group techniques typically proceed by introducing a field renormalization constant $\phi_0 = Z^{1/2}\phi_R$ and a mass counterterm to absorb ultraviolet divergences~\cite{peskin2018introduction}. In this conventional formalism, the anomalous dimension $\eta$ appears implicitly through the scale dependence of $Z$, which can obscure the direct physical link between the fixed-point action and the critical correlation function. 

To provide a more transparent derivation of the critical scaling, we employ a modified renormalization scheme that explicitly incorporates the anomalous dimension into the structure of the renormalized action.  This ``self-consistent" or ``bootstrap" approach avoids the intermediate calculation of $Z$-factors by directly enforcing scale invariance on the effective propagator. We introduce an arbitrary renormalization momentum scale $\mu$ to render the parameter dimensionless and partition the bare action into a scale-invariant fixed-point part and a set of counterterms:
\begin{align}
       \mathcal{S} &=  \frac{1}{2}\int\frac{d^dk}{(2\pi)^d}(\mu ^{2-\eta} r+|k|^{2-\eta})\phi_\mathbf{k}\phi_\mathbf{-k} \notag \\
       &+u\mu^{4-2\eta -d}\int\frac{d^dk_1d^dk_2d^dk_3}{(2\pi)^{3d}}\phi_\mathbf{k_1}\phi_\mathbf{k_2}\phi_\mathbf{k_3}\phi_\mathbf{-k_1-k_2-k_3}\notag\\
       &+ \frac{1}{2}\int \frac{d^dk}{2\pi}\left [\delta r\mu ^{2-\eta} +D(k)\right]\phi_\mathbf{k}\phi_\mathbf{-k}      \label{eq:renormalized action}\\
       &+\delta u\mu^{4-2\eta -d}\int\frac{d^dk_1d^dk_2d^dk_3}{(2\pi)^{3d}}\phi_\mathbf{k_1}\phi_\mathbf{k_2}\phi_\mathbf{k_3}\phi_\mathbf{-k_1-k_2-k_3}. \notag
\end{align}

In this decomposition, the non-interacting propagator is defined as $G_0(k) = (r + |k|^{2-\eta})^{-1}$. This form is chosen to explicitly contain the physical scaling behavior expected at the Wilson-Fisher fixed point, where the anomalous dimension $\eta$ is assumed to be a small parameter of order $O(\epsilon)$. The deviation of the bare kinetic term from this critical form is treated as a kinetic counterterm, $D(k) = |k|^\sigma - |k|^{2-\eta}$. The central logic of this method lies in the determination of $\eta$. Unlike standard schemes where counterterms are fixed by arbitrary subtraction conditions, here $\eta$ is determined via a self-consistency condition. Physical scale invariance at the critical point requires that the full propagator maintains the power-law form $G(k) \sim |k|^{-2+\eta}$ in the infrared limit. However, loop corrections to the self-energy, $\Sigma(k)$, generically introduce logarithmic terms $\sim \ln |k/\mu|$ that violate this scaling. The kinetic counterterm $D(k)$ is therefore tuned to exactly cancel these logarithmically divergent contributions. By expanding the long-range exponent $\sigma = 2 - \delta$, the kinetic counterterm takes the form:
\begin{equation}
D(k) = |k|^{2-\eta} \left[ (\eta-\delta)\ln|k| + O((\eta-\delta)^2) \right].
\label{eq:kinetic_counterterm}
\end{equation}
The coefficient $(\eta-\delta)$ serves as the ``counterterm" that must cancel the logarithmic momentum dependence arising from loop diagrams.

We first determine the interaction fixed point by imposing the renormalization condition on the four-point vertex $\Gamma_4$ at a symmetric momentum configuration defined at the scale $\mu$, where $|k_1+k_2|=|k_1-k_3|=|k_1-k_4|=\mu$:
\begin{align}
    \Gamma_4(k_1,k_2,k_3,k_4) \bigg|_{\text{sym}, \mu} = u\mu^{4-2\eta -d}.
\end{align}
Calculating the one-loop bubble diagram (Fig.~\ref{fig:diag-bubble}) in the massless limit ($r=0$) yields the vertex correction:
\begin{align}
    \Gamma_4|_\mu  = & (u+\delta u)\mu^{4-2\eta-d} + 4(n+8)\mu^{8-4\eta-2d}u^2  \notag \\
    &\times \int \left .\frac{d^dp}{(2\pi)^d}\frac{1}{|p|^{2-\eta}}\frac{1}{|k+p|^{2-\eta}}\right|_{k=\mu}.
\end{align}
Evaluating the integral and imposing the renormalization condition allows us to isolate the interaction counterterm:
\begin{equation}
    \delta u = 4(n+8)u^2 \frac{1}{(4\pi)^{d/2}}\frac{\Gamma(2-\eta-\frac{d}{2})\Gamma^2(\frac{d-2+\eta}{2})}{\Gamma(d-2+\eta)\Gamma^2(1-\frac{\eta}{2})}.
\end{equation}
Performing the $4-\epsilon$ expansion to extract the pole structure, we find the leading divergence:
\begin{equation}
    \delta u \approx \frac{n+8}{2\pi^2}u^2\frac{1}{\epsilon-2\eta}.
\end{equation}
The beta function is derived by demanding that the bare coupling is independent of the renormalization scale $\mu$, which is defined as the Callan-Symanzik (CS) equation~\cite{PhysRevD.2.1541,symanzik_small_1970}. This yields:
\begin{equation}
    \beta(u) \equiv  \mu\frac{\partial u}{\partial \mu} = -(\epsilon - 2\eta)u+ \frac{n+8}{2\pi^2}u^2.
\end{equation}
The non-trivial Wilson-Fisher fixed point $u^*$ is determined by the condition $\beta(u^*) = 0$, giving $u^* = \frac{2\pi^2}{n+8}(\epsilon - 2\eta)$. 

With the fixed-point coupling established, we proceed to the calculation of the anomalous dimension via the self-energy ``bootstrap". We compute the two-loop sunset diagram (Fig.~\ref{fig:diag-sunset}), which provides the leading momentum-dependent correction to the self-energy $\Sigma(k)$. The diagram evaluates to:
\begin{align}
    \Sigma^{S}(k) =& -16(n+2)u^2
    \int \frac{d^dk_1d^dk_2}{(2\pi)^{2d}} \notag\\
    &\times \frac{1}{|k_1|^{2-\eta}}\frac{1}{|k_2|^{2-\eta}}\frac{1}{|k_1+k_2+k|^{2-\eta}}  \\
    =& -\frac{16(n+2)u^2}{(4\pi)^d}|k|^{2d-6+3\eta}\frac{\Gamma^3(\frac{d-2+\eta}{2})\Gamma(\frac{6-3\eta-2d}{2})}{\Gamma^3(\frac{2-\eta}{2})\Gamma(\frac{3d-6+3\eta}{2})}\notag.
\end{align}
Expanding this result for small $\epsilon$ and $\eta$, and isolating the momentum-dependent part, we find a logarithmic violation of scaling:
\begin{equation}
    \Sigma^s = \frac{16(n+2)u^2}{(4\pi)^d}\frac{2}{2\epsilon-3\eta}|k|^{2-\eta} (1+2(2\eta-\epsilon)\ln|k|).
\end{equation} 
To restore the scale invariance of the theory at the fixed point, the coefficient of the $\ln|k|$ term in the self-energy must be exactly cancelled by the logarithmic term in the kinetic counterterm $D(k)$ from Eq.~\eqref{eq:kinetic_counterterm}. This self-consistency condition imposes:
\begin{equation}
    \eta - \delta =  \frac{(n+2)}{(n+8)^2}\frac{(\epsilon-2\eta)^3}{(2\epsilon-3\eta)}.
\end{equation}
We solve this equation iteratively by defining $\tilde{\eta} = \eta - \delta$. Keeping terms to the lowest consistent order in the expansion, we obtain the explicit expression
\begin{equation}
    \eta = \delta + \frac{(\epsilon-2\delta)^3}{2\epsilon -3\delta}\frac{n+2}{(n+8)^2}.
\end{equation}
This result confirms that $\eta$ is of order $O(\epsilon)$, consistent with our initial perturbative assumption. This derivation highlights the utility of the modified scheme: by directly enforcing the cancellation of logarithmic scaling violations, we determine the critical exponents naturally from the structure of the self-energy. 

We can immediately derive the correction to the scaling exponent $y_u$, which governs the approach to the fixed point, by evaluating the slope of the beta function at $u^*$:
\begin{equation}
y_u = -\frac{\partial \beta}{\partial u}\bigg|_{u^*} = -(\epsilon - 2\delta) +O(\epsilon^2). 
\end{equation} 
This negative exponent confirms the infrared (IR) stability of the fixed point in the critical plane.

Having determined the fixed-point coupling $u^*$ and the anomalous dimension $\eta$, we proceed to calculate the critical exponent $\nu$, which characterizes the divergence of the correlation length near the critical temperature, $\xi \propto |T-T_c|^{-\nu}$. In the renormalization group framework, $\nu$ is the inverse of the eigenvalue $y_t$ associated with the mass operator ($\nu = 1/y_t$). Physically, $y_t$ quantifies how strongly the mass parameter $r$ scales under the renormalization group transformation. It is derived from the beta function $\beta_r$ by linearizing the flow around the fixed point.

To determine $\beta_r$, we analyze the renormalization of the mass term using the condition that the self-energy at zero momentum equals the renormalized mass: $\Sigma(k=0) = r\mu^{2-\eta}$. The leading-order correction arises from the one-loop tadpole diagram (Fig.~\ref{fig:diag-tadpole}) as:
\begin{equation}
    \Sigma(0) = 2(n+2)u\mu^{4-2\eta-d}\int \frac{d^dp}{(2\pi)^d}\frac{1}{r\mu^{2-\eta}+|p|^{2-\eta}} + \delta r \mu^{2-\eta}.
\end{equation}
By utilizing the RG conditions and tuning the counterterm to cancel the divergence in the integral, we obtain:
\begin{equation}
    \delta r =  \frac{n+2}{2}u  \frac{r}{\pi^2}\frac{1}{\epsilon-2\eta}.
\end{equation}
After solving the CS equation for the invariance of the bare mass, one will get the beta function as 
\begin{equation}
    \beta_r = -(2-\eta) r + r\frac{n+2}{2\pi^2}u.
\end{equation}
The thermal eigenvalue $y_t$ is defined as the negative scaling dimension of the mass at the fixed point:
\begin{equation}
    y_t =- \frac{d\beta_r}{dr}\bigg|_{u^*} = 2-\delta -\frac{n+2}{n+8}(\epsilon-2\delta)+O(\epsilon^2).
\end{equation}
This result establishes the correlation length exponent $\nu = y_t^{-1}$, completing our description of the critical universality class to leading order.

\subsection{$Z$-factor renormalization scheme}
To validate the results derived via the self-consistent bootstrap method, we also perform the calculation using the standard field-theoretic renormalization group (RG) formalism. In this scheme, we introduce explicit renormalization constants to absorb divergences, maintaining the canonical form of the propagator $G^{-1} \sim |k|^\sigma$. The renormalized action is given by:
\begin{align}
       \mathcal{S} =& \frac{1}{2}\int \frac{d^dk}{(2\pi)^d}(\mu^{\sigma}
       r+|k|^\sigma)\phi_\mathbf{k}\phi_\mathbf{-k} \notag\\
       &+\mu^{2\sigma-d}u\int\frac{d^dk_1d^dk_2d^dk_3}{(2\pi)^{3d}}\phi_\mathbf{k_1}\phi_\mathbf{k_2}\phi_\mathbf{k_3}\phi_\mathbf{-k_1-k_2-k_3} \notag\\
       &+ \frac{1}{2}\int \frac{d^dk}{(2\pi)^d}(\delta r\mu^{\sigma}+\delta Z|k|^\sigma)\phi_\mathbf{k}\phi_\mathbf{-k}  \\
       &+\delta u\mu^{2\sigma-d}\int\frac{d^dk_1d^dk_2d^dk_3}{(2\pi)^{3d}}\phi_\mathbf{k_1}\phi_\mathbf{k_2}\phi_\mathbf{k_3}\phi_\mathbf{-k_1-k_2-k_3}\notag. 
\end{align}
Inspired by Ref.~\cite{blanchard2013}, we defined the RG conditions as 
\begin{align}
    \Sigma(k = 0) &= r,\label{eq:RG_conditions1}\\
     \frac{d\Sigma}{dk^{\sigma}} \bigg|_{k=\mu} &= 0,\label{eq:RG_conditions2}\\
     \Gamma_4(k_1,k_2,k_3,k_4)\bigg|_{\text{sym}, \mu} &=u \label{eq:RG_conditions3}. 
\end{align}
Similarly, we can obtain the leading order of the beta function as $\beta_u =-(\epsilon-2\delta)u +\frac{n+8}{2\pi^2}u^2$, where we have the fixed point $u^*$ as $u^* = \frac{2\pi^2}{n+8}(\epsilon-2\delta)$, with $y_u = -(\epsilon-2\delta)+O(\epsilon^2)$ by using the the RG conditions Eq.~\eqref{eq:RG_conditions3}, calculating the bubble diagram in Fig.~\ref{fig:diag-bubble} and solving the CS equation. For the mass renormalization, we can also handle it with the same approach for Fig.~\ref{fig:diag-tadpole} and get 
$y_t = 2-\delta -\frac{n+2}{n+8}(\epsilon-2\delta)+O(\epsilon^2)$.

For the anomalous dimension for the critical Green's function $\eta$, we calculate the first momentum-dependence self-energy diagram in Fig.~\ref{fig:diag-sunset} as  
\begin{align}
    \Sigma^{(2s)} =& -16(n+2)u^2\mu^{4\sigma-2d}\int \frac{d^dk_1d^dk_2}{(2\pi)^{2d}} \notag \\
    &\times\frac{1}{|k_1|^\sigma}\frac{1}{|k_2|^\sigma}\frac{1}{|k_1+k_2+k|^\sigma}\\
        =& -\frac{(n+2)u^2}{16\pi^4}\frac{\mu^{4\sigma-2d}}{(|k|^\sigma)^{(3\sigma-2d)/\sigma}}\frac{\Gamma^3(\frac{d-\sigma}{2})\Gamma(\frac{3\sigma-2d}{2})}{\Gamma^3(\frac{\sigma}{2})\Gamma(\frac{3d-3\sigma}{2})} \notag.
\end{align}
By considering the renormalization conditions in Eq.~\eqref{eq:RG_conditions2} and performing the $(\delta,\epsilon)$ expansion, we have 
\begin{align}
    \delta Z & = - \frac{\partial \Sigma^{(2s)}}{\partial|k|^\sigma} \bigg|_{k=\mu} \notag\\
    &=\frac{(n+2)u^2}{16\pi^4}\frac{3\sigma-2d}{\sigma}\frac{\Gamma^3(\frac{d-\sigma}{2})\Gamma(\frac{3\sigma-2d}{2})}{\Gamma^3(\frac{\sigma}{2})\Gamma(\frac{3d-3\sigma}{2})} \notag\\
    & =  \frac{(n+2)u^2}{16\pi^4}\frac{2}{\sigma}\frac{\Gamma^3(\frac{d-\sigma}{2})\Gamma(\frac{3\sigma-2d
    }{2}+1)}{\Gamma^3(\frac{\sigma}{2})\Gamma(\frac{3d-3\sigma}{2})} \label{Gammafunc_struc}\\
        & =  \frac{(n+2)u^2}{16\pi^4}\frac{\Gamma^3(1-\frac{\epsilon-\delta}{2})\Gamma(\frac{2\epsilon-3\delta}{2})}{\Gamma^3(1-\frac{\delta}{2})\Gamma(3-\frac{3\epsilon-3\delta}{2})} \notag\\
        &\approx \frac{(n+2)u^2}{8\pi^4}\frac{1}{2\epsilon-3\delta}.
\end{align}
We then solve the CS equation for the two-point correlation function, which states that the two-point correlation function of the bare theory remains unchanged for any momentum scale $\mu$
\begin{equation}
    \left[\mu\frac{\partial}{\partial \mu} + \beta(u)\frac{\partial}{\partial u} + 2\gamma(u)\right] G^{(2)}(k) =0 \text{   for }\forall k,  
\end{equation}
where $\gamma$ is the anomalous dimension for the field $\phi$.  Therefore, we have 
\begin{equation}
    \gamma(u) = -\frac{1}{2}\beta(u) \frac{\partial \delta Z }{\partial u}  =  \frac{(n+2)u^2}{8\pi^4}\frac{\epsilon-2\delta}{2\epsilon-3\delta}.
\end{equation}
Therefore, at the fixed point $u=u^*$, we have 
\begin{equation}
    \eta \equiv 2-\sigma + 2\gamma(u^*) = \delta + \frac{n+2}{(n+8)^2}\frac{(\epsilon-2\delta)^3}{2\epsilon-3\delta}. 
\end{equation}
This result is fully consistent with the value derived via the bootstrap method in the previous section, confirming the robustness of our 4-$\epsilon$ expansion. 

It is worth noting that our renormalization group analysis, firmly established in the $d=4-\epsilon$ framework, resolves a subtlety often overlooked in standard long-range RG treatments. Conventional approaches typically restrict their analysis to the immediate vicinity of the mean-field limit (where $d \approx 2\sigma$), treating the long-range exponent $\sigma$ as a fixed parameter. Under this restricted view, the Gamma function term $\Gamma(\frac{3\sigma-2d}{2}+1)$ in Eq.~\eqref{Gammafunc_struc} appears convergent for generic values of $\sigma < 2$, leading to the conclusion that $\delta Z = 0$ and consequently $\eta = 2-\sigma$ (i.e., no anomalous correction). However, this argument fails near the short-range crossover $\sigma \to 2$, where the Gamma function develops a pole structure $\sim \frac{1}{2\epsilon - 3\delta}$. In our $d=4-\epsilon$ framework, considering the non-mean-field regime naturally implies that $\delta = 2-\sigma$ acts as a small quantity of order $O(\epsilon)$. By strictly treating $\delta$ as an $O(\epsilon)$ term, our method explicitly retains the denominator $2\epsilon - 3\delta$ as a small quantity within the perturbative expansion, thereby capturing the singularity that drives the non-trivial correction to the anomalous dimension. We note that Ref.~\cite{blanchard2013} also obtained a second-order correction for $\eta$.
However, their result relies on the assumption $\delta \ll \epsilon' = 2\sigma - d$, whereas our method treats $\delta$ as being of the order of $\epsilon$. Consequently, due to these different expansion schemes, their derived result is distinct from ours.


\section{Discussions}

We have established a comprehensive picture of the renormalization group flow governing the competition between long-range and short-range interactions. By performing a rigorous field-theoretical RG analysis up to the two-loop level within a controlled $\epsilon$-expansion ($d=4-\epsilon$), we successfully capture the singular behavior inherent to the crossover region (specifically near $\sigma \to 2$). The success of this framework stems from the fact that, in the $d=4-\epsilon$ dimension, the non-mean-field regime naturally implies that $\delta = 2-\sigma$ is of order $\mathcal{O}(\epsilon)$; this ensures that our perturbative expansion remains well-defined throughout the entire crossover region, resolving the singularities that plagued previous single-parameter expansions. We validated our results via a novel perturbative bootstrap scheme. By explicitly embedding the anomalous dimension into the action to enforce scale invariance, this method bypasses the ambiguity of implicit $Z$-factors and transparently captures the singular structure near the crossover.

Our central finding reveals a non-trivial correction to the anomalous dimension $\eta$ beyond the mean-field level, suggesting that the foundational assumption of Sak's criterion (that $\eta$ remains fixed until it intersects the short-range value) is invalid. Consequently, our results provide strong theoretical support for the alternative scenario where the crossover threshold is strictly located at the geometric value $\sigma_* = 2$, consistent with the recent high-precision numerical results~\cite{xiao_saks_2026,picco_critical_2012,blanchard2013,Xiao2025_XY,yaodingyun2025,liu_two-dimensional_2025,Grassberger_2013}.


Looking forward, several avenues for future research remain. First, we plan to extend the current framework to perform multi-loop calculations. Obtaining higher-order results for the critical exponents is essential for a quantitative comparison with the high-precision Monte Carlo data now available. Second, we aim to develop a more refined field-theoretical description that explicitly renormalizes the short-range operator $K_s(-\Delta)$ alongside the long-range term. This would allow us to quantitatively derive the RG flow exponent $y_{K_s}$ and establish a precise flow equation for the global RG diagram (Fig.~\ref{fig:RG-flow}), fully mapping the trajectory from the long-range to the short-range fixed point. 

Our perturbative calculations, albeit quantitatively valid only in $(4-\epsilon)$ dimensions, 
strongly suggest that, in low spatial dimensions, a consistent theoretical understanding  
of classical and quantum systems with long-range interactions is far from being established.
In two dimensions, there is a host of exact critical exponents for systems with short-ranged interactions, thanks to Coulomb-gas arguments, conformal field theory and  stochastic Loewner evolution theory, while, in contrast, scarce information is available for long-range interacting systems; actually, high-precision numerical studies had remained preliminary until very recently. 
In three dimensions, the situation is even worse, where exact critical exponents are nearly absent even for the short-range case. 
For one-dimensional systems with long-range interactions, 
finite-temperature phase transitions can occur only if decay exponent $\sigma \leq 1$.
It is conjectured~\cite{xiao2026universality} that the local conformal invariance 
is ill-defined for $\sigma < 1$, since the diffusion in the L\'evy flights becomes hyper-ballistic. An adjoint hypothesis is that, 
in the non-classical regime with $(d/2 < \sigma < 1, d<2)$, the anomalous dimension 
is locked at the mean-field value $\eta = 2-\sigma$ while the correlation-length 
exponent $\nu$ takes non-trivial value. 
Given many unresolved issues, 
we expect that this work would spur research interest in 
long-range interacting systems, both theoretically and numerically.

\section*{Acknowledgement} The authors express gratitude to Zhijie Fan, Sheng Fang and Pengcheng Hou for their engaging and valuable discussions. K. C. was supported by the National Key Research and Development Program of China, Grant No. 2024YFA1408604, the National Natural Science Foundation of China under Grants No. 12474245 and
No. 12447103, and the GHfund A(202407010637). Z.L, P.H., and Y.D. were supported by the National Natural Science Foundation of China (under Grant No. 12275263), the Innovation Program for Quantum Science and Technology (under grant No. 2021ZD0301900), the Natural Science Foundation of Fujian Province of China (under Grant No. 2023J02032).  

\bibliography{Ref}

\end{document}